\documentclass{raa} 

\usepackage{graphicx,times}             
\usepackage{natbib}
\usepackage{amssymb,amsmath}
\bibpunct{(}{)}{;}{a}{}{,}

\usepackage[pagebackref=true]{hyperref}

\begin{document}

  \title{Variable stars detection in the field of open cluster NGC 188
}

   \volnopage{Vol.0 (20xx) No.0, 000--000}      
   \setcounter{page}{1}          

   \author{Fang-Fang Song 
      \inst{1,2}
   \and Hu-Biao Niu
      \inst{1,2}
   \and Ali Esamdin
      \inst{1,2}
   \and Yu Zhang
      \inst{1,2}
   \and Xiang-Yun Zeng
      \inst{3}

   }

   \institute{Xinjiang Astronomical Observatory, Chinese Academy of Sciences,
             Urumqi, Xinjiang 830011, People's Republic of China; {\it niuhubiao@xao.ac.cn}\\
        \and
             School of Astronomy and Space Science, University of Chinese Academy of Sciences, Beijing 	            100049, People's Republic of China;\\
        \and
             Center for Astronomy and Space Sciences, China Three Gorges University, Yichang 443000, People's Republic of China.\\
\vs\no
   {\small Received 20xx month day; accepted 20xx month day}}

\abstract{ This work presents the charge-coupled device (CCD) photometric survey of the old open cluster NGC 188. Time-series V-band photometric observations were conducted for ten nights in January 2017 using the Nanshan One-meter Wide-field Telescope (NOWT) to search for variable stars in the field of the cluster field. A total of 25 variable stars, including one new variable star, were detected in the target field. Among the detected variables, 16 are cluster member stars, and the others are identified as field stars. The periods, radial velocities, effective temperatures, and classifications of the detected variables are discussed in this work. Most of the stars' effective temperatures are between 4200 K and 6600 K, indicating their spectral types are G or K.  The newly discovered variable is probably a W UMa system. In this study, a known cluster variable star (V21 = V0769 Cep) is classified as an EA-type variable star based on the presence of an 0.5 magnitude eclipse in its light curve.
\keywords{Galaxy --- open cluster: individual: NGC 188 --- stars: variables: general --- technique: photometric --- method: data analysis}
}

   \authorrunning{F.-F. Song, H.-B. Niu, et al }            
   \titlerunning{Variable Stars in open cluster NGC 188 }  

   \maketitle

%
%
\section{Introduction}           
\label{sect:intro}

Open Clusters (OCs), which are composed of young stars, are crucial for studying the formation and evolution of the stellars and Galactic disk \citep{2006A&A...445..545P, 2014NewA...32...36O, 2020MNRAS.495.1531G}. Because the cluster members born from the same interstellar cloud are assumed to have a common age, distance, reddening, and chemical abundance, the studies in cluster variables provide significant clues to probe the structure and evolution of stars and clusters. The high precision in the astrometric and photometric measurements by Gaia gives more opportunities to classify the cluster member stars from the field stars, which is great progress for the studies of variable stars in the open clusters \citep{2016A&A...595A...1G, 2020A&A...640A...1C}.   

As one of the most ancient, rich open clusters known in our Milky Way, NGC 188 ( l = 122.843 $\deg$, b = + 22.384 $\deg$; C 0039+850) is a captivating open cluster that is intensively studied by numerous studies \citep{1999AJ....118.2894S, 2010AJ....139.1942F, 2015AJ....149...94H, 2020AJ....159...11C}. It is an excellent laboratory based on its abundant member stars, easy to observe, and less contaminated by field stars owning to its special location that it is Located at high latitude and far away from the galactic disk \citep{2005A&A...433..917B, 2007AJ....133.1409F, 2015AJ....150...61W}. 857 cluster member stars with probabilities over 70\% was identified by \citet{2020A&A...640A...1C}. Lists of the previous observational surveys for the cluster are shown in \cite{2007AJ....133.1409F, 2008AJ....135.2264G, 2015AJ....150...61W}. Various works about the fundamental parameters of NGC 188 have been performed after the release of the Gaia data which provide the unprecedented precise parallax measurements \citep{2020A&A...640A...1C, 2020MNRAS.499.1874M, 2019MNRAS.483.2758B, 2018A&A...616A..10G}. Taken the fitted parameters by \cite{2020A&A...640A...1C}, The basic cluster parameters of NGC 188 are as follows, the distance modulus 11.15 mag, corresponding to the distance 1698 pc, $log(Age) = 9.85 yr$, extinction $A_V = 0.21 mag$. 

Also, NGC 188 is a special cluster owning for its abundant and various variable stars. Early in the 1960s, four short-period W UMa stars (known as EQ Cep, ER Cep, ES Cep, EP Cep) and a suspected variable NSV 395 were discovered by \cite{1964AN....288...49H}. \cite{1987ApJ...314..585K} and \cite{1990AcA....40...61K} discovered another 7 short-period variables through CCD photometric surveys using the 0.9 m telescope at Kitt Peak National Observatory (KPNO). Subsequently, \cite{2002ChJAA...2..481X} and \cite{2004MNRAS.355.1369Z} yielded sixteen variables by monitoring the cluster in 1 $deg^2$ field with the 60/90-cm Schmidt telescope located at the Xinglong Station of the National Astronomical Observatories of the Chinese Academy of Sciences. Meanwhile, \cite{2003AJ....126..276K} reported 51 faint variables in the cluster's central area of $17 \times 17~arcmin^2$ with the WIYN 3.5-m telescope, but only two variables were detected by the subsequent monitoring of other telescopes. Another 18 variables were discovered by \cite{2008AcA....58..263M} as the results of searching for transiting planets in the open cluster NGC 188, part of the project of Planets in Stellar Clusters Extensive Search (PISCES). In the field of $1.5 \times 1.5~deg^2$ around the cluster, 18 new variable stars were identified by the MASTER series robotic telescope in the Astronomical Observatory of Ural Federal University \citep{2013OEJV..157....1P}. These variable stars were included in the electronic catalog of the American Association of Variable Star Observers (AAVSO/VSX\footnote {http://www.aavso.org/vsx/}) in which 68 variable stars are known within a 30 arcmin radius around NGC 188. The Gaia collaboration identified another 61 variable stars based on the the Gaia DR3 database \citep{2022yCat.1358....0G}. A total of 129 variables were collected in this field.

This paper is structured as follows. In Section 2 the observations and data reductions are presented. We focused on the variable identification and memberships in Section 3.
In Section 4 we compared our results with previous works and emphasized our work of the new variable star. The conclusions are given in Section 5.


\section{Observations and Data Reductions}
\label{sect:Obs}

\begin{figure}
\includegraphics[width=\linewidth]{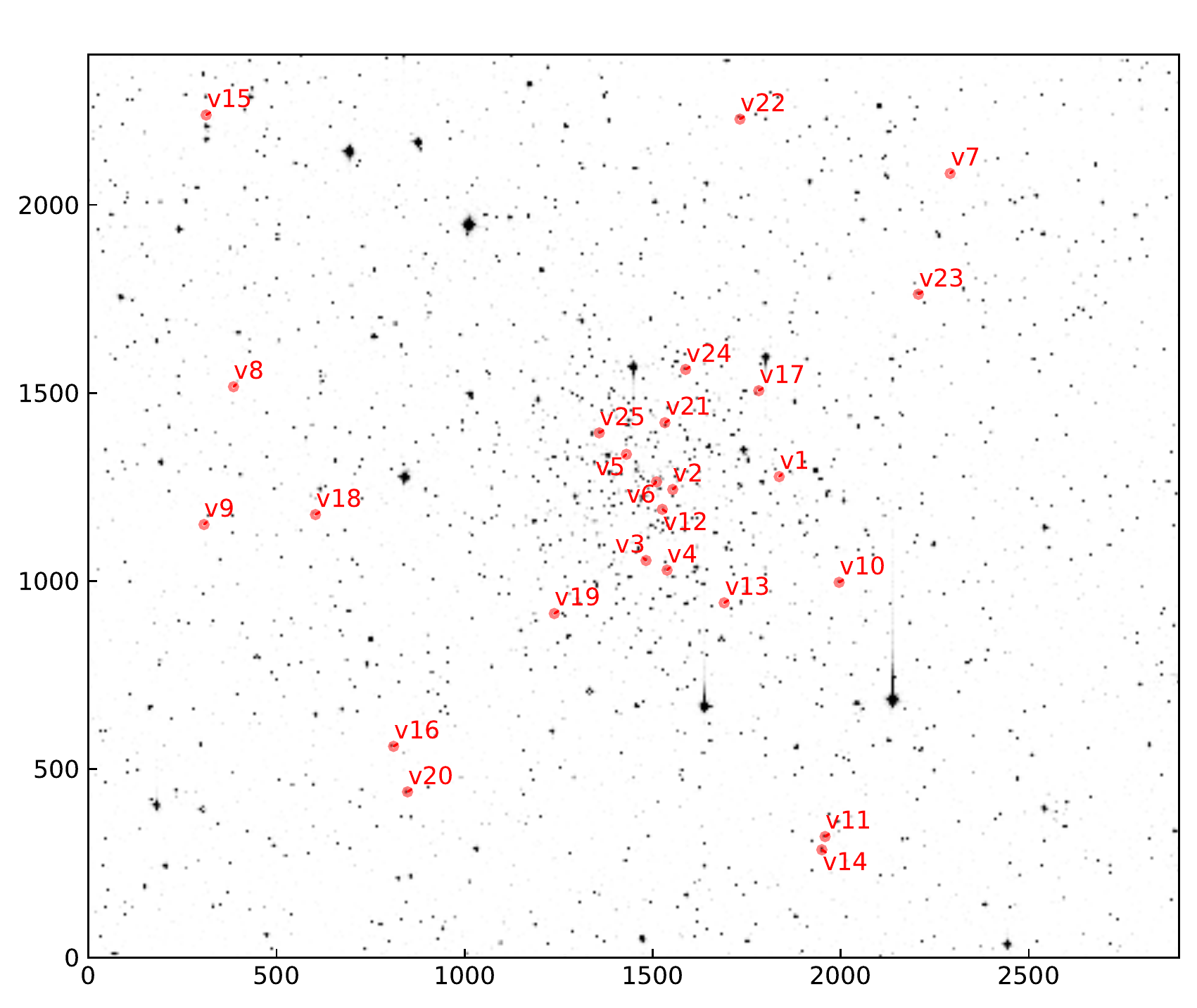}
\caption{The field of NGC 188 observed using the Nanshan One-meter Wide-field Telescope (NOWT) with the detected variable stars marked in red circles. North is up and east is to the left. }
\label{fig:chart}
\end{figure}

In January 2017, ten days of photometric observations were conducted by using the Nanshan One-meter Wide-field Telescope (NOWT; \cite{2016RAA....16..154S, 2018Ap&SS.363...68M, 2020RAA....20..211B}) located at the Nanshan station of Xinjiang Astronomical Observatory (XAO), Chinese Academy of Sciences (CAS). An E2V CCD203-82 (blue) chip CCD camera with 4096 $\times$ 4136 pixels was mounted at the prime focus of the telescope, providing a field of view (FOV) of 1.3 $\times$ 1.3 $deg^{2}$. The telescope was equipped with a Johnson-Cousins standard UBVIR filter system for broadband photometry and operated at -120$^{\circ}$C with liquid nitrogen cooling. During the observations, a number of bias and twilight flat frames were taken, and the seeing in all of these images was below 2.2 arcsec. In total, over 79 hours with 4198 images of useful data were obtained on 10 nights. The journal of the observations of NGC 188 is listed in Table \ref{tab:obs}. The exposure time of time-series observations for the V band is 35s. And the typical field of view in this period is 54.62 $\times$ 45.2 $arcmin^2$, corresponding to 2900 $\times$ 2400 pixels. The observed field of NGC 188 is shown in Figure \ref{fig:chart}.

\begin{table}
\centering
\fontsize{8} {10pt}\selectfont
\tabcolsep 0.10truecm
\caption{Journal of CCD photometric observations for NGC 188 where N represents the number of images and Exp represents observation exposure time for each filter.}
\begin{tabular}{rccrrrrrrrrrrrrrr}
\hline\hline
\multicolumn{1}{c}{UT date}         &
\multicolumn{1}{c}{Object}       &
\multicolumn{1}{c}{Pixel}       &
\multicolumn{1}{c}{Duration}         &
\multicolumn{1}{c}{V}             \\
\multicolumn{1}{c}{}                &
\multicolumn{1}{c}{}                &
\multicolumn{1}{c}{}                &
\multicolumn{1}{c}{(Hour)} &
\multicolumn{1}{c}{(N$\times$Exp)} \\
\hline
05 Jan 2017  & NGC 188 & 2900$\times$2400 & 1.2     & 38  $\times$ 35s   \\		 
06 Jan 2017  & NGC 188 & 2900$\times$2400 & 1.8     & 58  $\times$ 35s   \\
07 Jan 2017  & NGC 188 & 4096$\times$4136 & 5.1     & 171 $\times$ 35s  \\
             &         & 2900$\times$2400 & 2.6     & 110 $\times$ 35s  \\							
08 Jan 2017  & NGC 188 & 2600$\times$2400 & 4.5     & 267 $\times$ 35s  \\
             &         & 2900$\times$2400 & 5.4     & 290 $\times$ 35s  \\
09 Jan 2017  & NGC 188 & 2900$\times$2400 & 12.5    & 649 $\times$ 35s \\
10 Jan 2017  & NGC 188 & 2900$\times$2400 & 12.5    & 657 $\times$ 35s \\
11 Jan 2017  & NGC 188 & 2900$\times$2400 & 11.0    & 598 $\times$ 35s \\
12 Jan 2017  & NGC 188 & 2900$\times$2400 & 0.7     &  67 $\times$ 35s  \\
13 Jan 2017  & NGC 188 & 2900$\times$2400 & 9.7     & 502 $\times$ 35s  \\
14 Jan 2017  & NGC 188 & 2900$\times$2400 & 12.5    & 666 $\times$ 35s  \\
\hline   
\end{tabular}
\label{tab:obs}
\end{table}

\begin{figure}
\includegraphics[width=\linewidth]{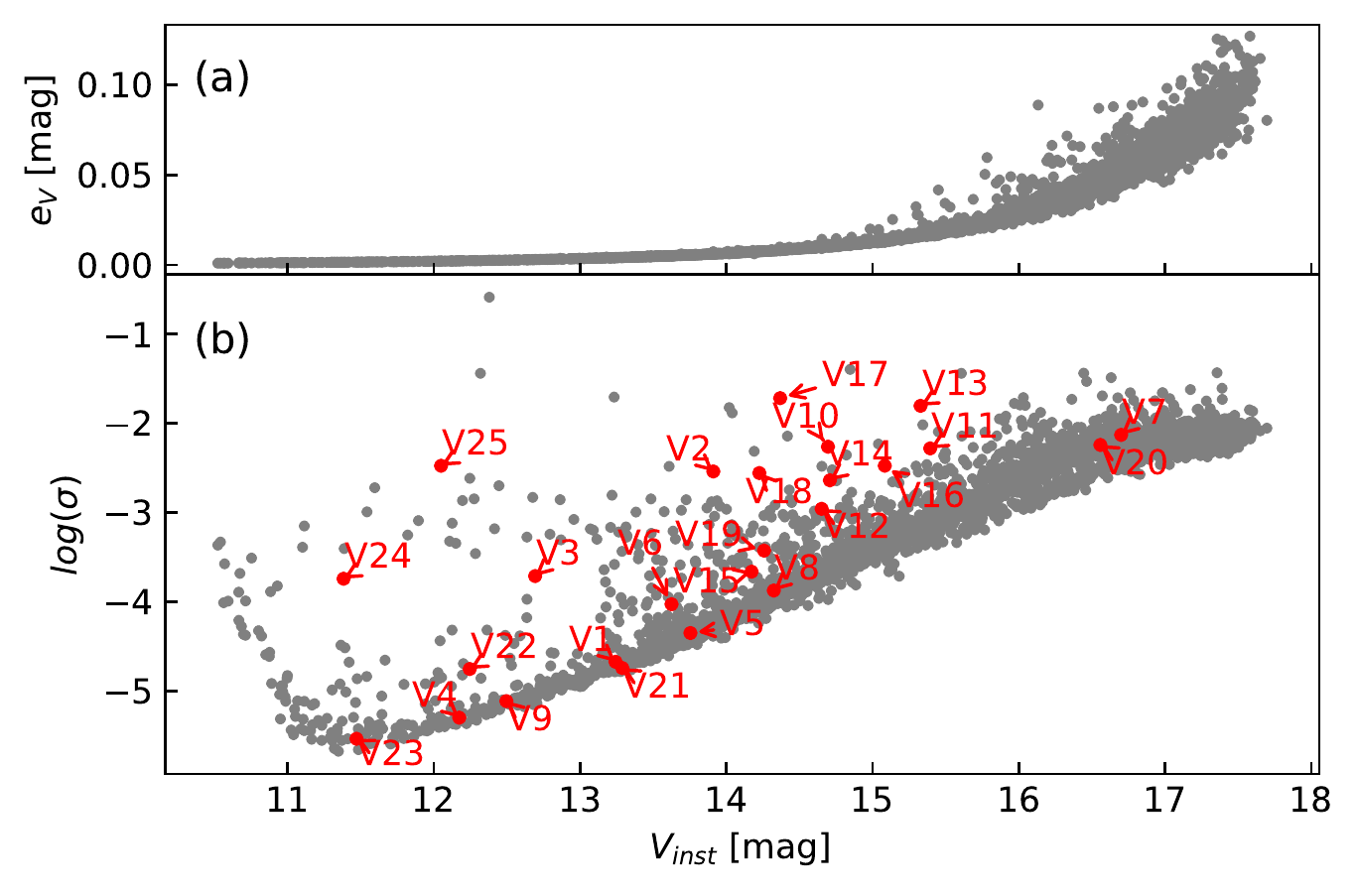}
\caption{Panel(a) represents the photometric errors of this work as a function of instrumental magnitude in the V band, meanwhile, Panel(b) is the standard deviation as a function of magnitude. The red dots represent the variables stars identified in this work.}
\label{fig:qua}
\end{figure}

The data reduction steps followed the standard procedures employed for optical CCD aperture photometry. The observed images were pre-processed by IRAF\footnote{Image Reduction and Analysis Facility, http://iraf.noao.edu/} for overscan and bias subtraction and division flat-fields correction. The dark correction was ignored because the telescope was operated in a low-temperature environment and the thermionic noise was less than 1 e $pix^{-1}h^{-1}$. The instrumental magnitudes of the stars in each frame were extracted by the automated photometric software SExtractor\citep{1996A&AS..117..393B} and the equatorial coordinates (RA, DEC) for each detected star were computed by triangular matching with the UCAC4 \citep{2013AJ....145...44Z}. Figure \ref{fig:qua} (a) represents the photometric errors of this work as a function of instrumental magnitudes in the V band, which indicates that the photometric errors are less than 0.1 mag when the observational magnitudes are brighter than 17 mag. Following the same procedure described in \cite{2016RAA....16..154S, 2018Ap&SS.363...68M, 2021RAA....21...68L}, the differential light curves of 3585 stars were obtained using the data processing system of our XAO time-domain survey pipeline. 

\section{Variables identification}
\label{sect:data}

\begin{figure*}
   \centering
   \includegraphics[width=\linewidth]{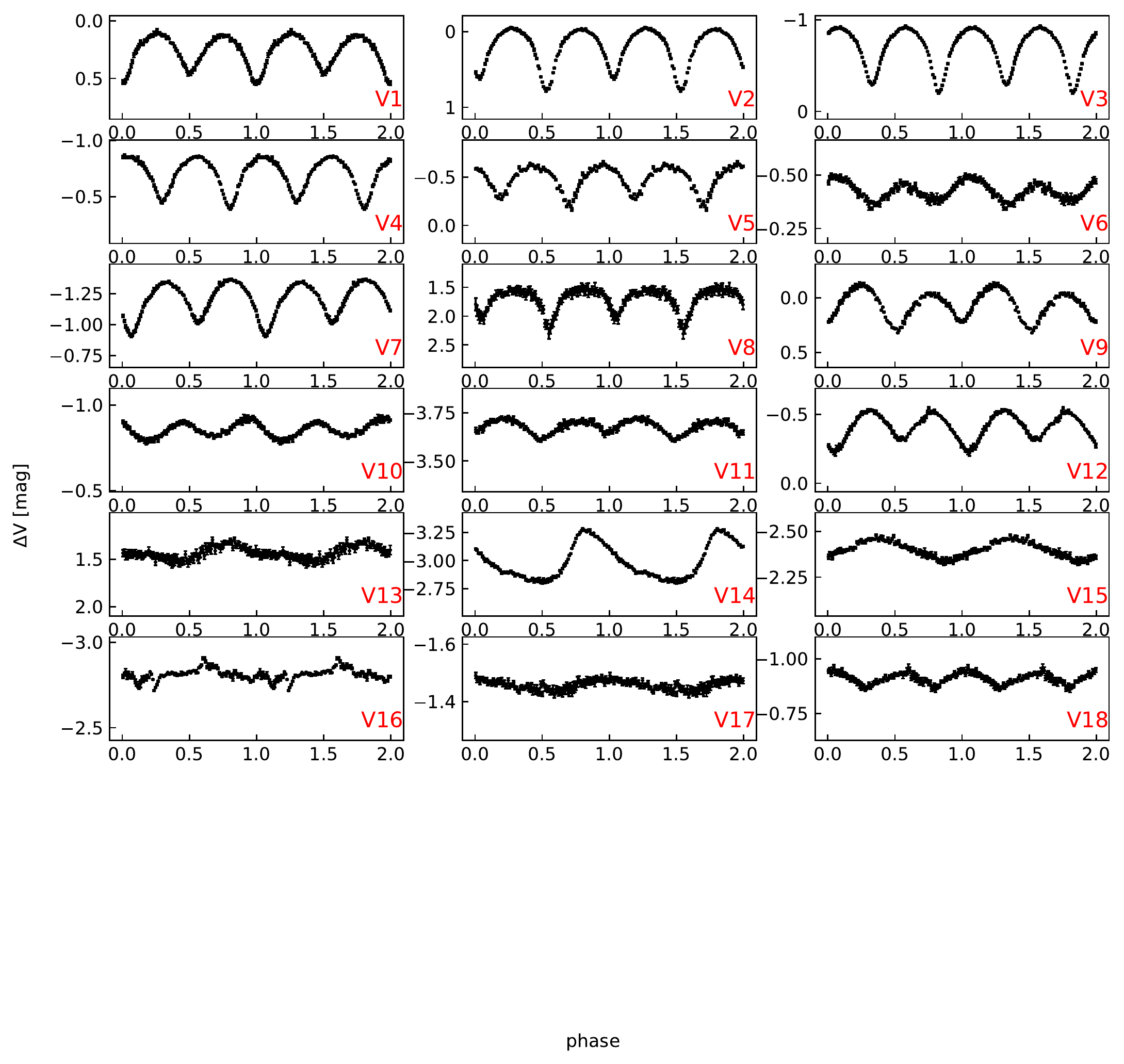}
   \caption{ The phased light curves of periodic variable stars.}
   \label{fig:fig1}
\end{figure*}

\begin{figure*}
   \centering
   \includegraphics[width=\linewidth]{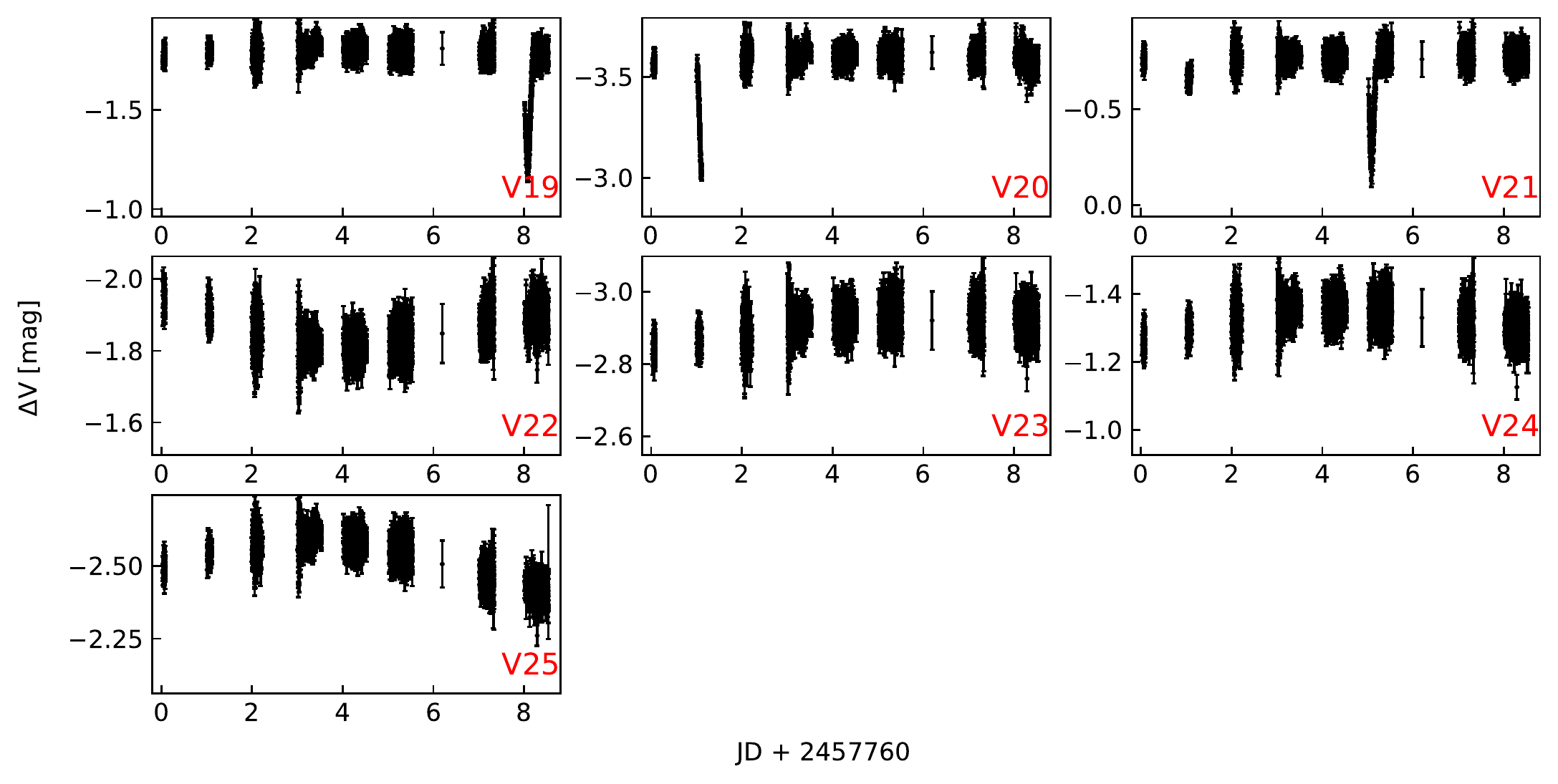}
   \caption{ The light curves of long-period variable stars.}
   \label{fig:fig2}
\end{figure*}

We examined the light curves of all detected stars by visual inspection and found 25 stars with obvious light variability.These variables were carefully examined for any blending or contaminating by neighboring stars. The main characteristics of the 25 variable stars are given in Table \ref{tab:vari}. The finding chart of these variable stars in the field of the cluster are shown in Figure \ref{fig:chart}. These stars are not at the edge of CCD frames, and their light curves do not have outliers. The light variations of these variables can be seen in Figures \ref{fig:fig1} and \ref{fig:fig2}. To investigate the spread in data points for variable stars, Figure \ref{fig:qua} (b) demonstrates the root-mean-square (RMS, labeled as $\sigma$) scatter as a function of instrumental V magnitude. It indicates that, in general, detected variable stars (the red dots) have larger standard deviation in magnitude compared to non-variable stars.

The periods of periodic variable stars were obtained by the Generalized Lomb-Scargle Periodogram (GLS; \cite{2009A&A...496..577Z}) considering the effect of noise as a sub-package of PyAstronomy \citep{2019ascl.soft06010C}. The generalised Lomb-Scargle periodogram computes the error-weighted Lomb-Scargle periodogram, and provides more accurate frequencies compared with the Lomb-Scargle periodogram. The adopted periods correspond to the frequency of the maximum power and the accuracies of those values are influenced by the mean magnitude, amplitude, and measurement errors. To avoid false variability arising solely due to the aliasing effect, we rechecked all the calculated periods and none of them can be seen as a factor of one day. After binning in intervals of 0.01 phase, we calculate the mean magnitude in each bin. The resulting phase-folded light curves of periodic variables are shown in Figures \ref{fig:fig1}. Our data are insufficient to yield accurate periods for the other 7 long-period variable stars, and the light curves of these variables are shown in Figure \ref{fig:fig2}.

For the amplitudes of the periodic variables, phased data were sorted from small to large first, then we averaged the sorted data with a sampling interval of 60 data points using the moving average method \citep{2022EPJP..137...50S}. Then the amplitude of the variation in periodic variables was calculated by substracting the minimum values from the maximum values. We did almost the same operation for the amplitudes of other detected variables, except that the data were sorted by Julian Dates. \cite{2019AJ....158...93B} gives the stellar effective temperature regression for the second Gaia data release applied the supervised machine-learning algorithm, based on the combination of the stars in four spectroscopic surveys: the Large Sky Area Multi-Object Fiber Spectroscopic Telescope, Sloan Extension for Galactic Understanding and Exploration, the Apache Point Observatory Galactic Evolution Experiment, and the Radial Velocity Extension. 
We have given the effective temperatures of most variables by cross-match the coordinates with \cite{2019AJ....158...93B}, except V11 and V15, which have no values of effective temperatures given in this table due to inconformity with source selection criteria of  \cite{2019AJ....158...93B}. The effective temperature of these variable stars ranges from 4200 to 6700K, corresponding to the spectral types of K - F types.

The identified variables were cross-checked with the International Variable Star Index (VSX\footnote {http://www.aavso.org/vsx/}) and the catalogue of Gaia DR3 Part 4 Variability \citep{2022yCat.1358....0G}, and found most variables are matched with the online catalogs except one star V18, which implies that in our sample only V18 is a new discovery. The periods, amplitudes and shapes of the light curves of known stars produced by this work are mostly consistent with those given on the VSX website. We found an about 0.5 mag eclipse appeared in the light curve of V21,  which implies that it could be an Algol-type eclipsing binary rather than a BY Draconis-type variable star given by \citet{2008AcA....58..263M}.

\begin{table*}
\centering
\fontsize{8} {10pt}\selectfont
\tabcolsep 0.10truecm
\caption{Basic Parameters of the Detected Variable stars in the field of the open cluster NGC 188}
\begin{tabular}{rccrrrrrrrrrrrrrr}
\hline\hline
\multicolumn{1}{c}{ID}         &
\multicolumn{1}{c}{Coords (J2000)}        &
\multicolumn{1}{c}{G}        &
\multicolumn{1}{c}{BP-RP}         &
\multicolumn{1}{c}{Distance}         &
\multicolumn{1}{c}{parallax}        &
\multicolumn{1}{c}{pmRA}         &
\multicolumn{1}{c}{pmDE}        &
\multicolumn{1}{c}{Amplitude}         &
\multicolumn{1}{c}{Period}        &
\multicolumn{1}{c}{Memb}        &
\multicolumn{1}{c}{$T_eff$}        &
\multicolumn{1}{c}{Type}          \\
\multicolumn{1}{c}{}                &
\multicolumn{1}{c}{(hh mm ss dd mm ss)}  &
\multicolumn{1}{c}{(mag)} &
\multicolumn{1}{c}{(mag)} &
\multicolumn{1}{c}{(pc)} &
\multicolumn{1}{c}{mas}        &
\multicolumn{1}{c}{mas/yr}         &
\multicolumn{1}{c}{mas/yr}        &
\multicolumn{1}{c}{(mag)} &
\multicolumn{1}{c}{(day)}           &
\multicolumn{1}{c}{(yes/no?)}           &
\multicolumn{1}{c}{(K)}           &
\multicolumn{1}{c}{}           \\
\hline
V1   & 00 46 54.53 +85 21 44.1 & 16.525(6) & 1.12(2) & $1587_{-87}^{+98}$  &0.58(4)   &-2.47(5)   & -0.84(4)  &0.43(5) & 0.28961(2)  &  yes & $5423 \pm 300$  & EW/KW  \\
V2   & 00 47 33.92 +85 16 24.8 & 16.49(1)  & 1.16(6) & $1568_{-86}^{+97}$ &0.61(4)   &-2.47(5)   & -0.76(4)  &0.83(6) & 0.30675(2)  &  yes & $5293 \pm 261$  & EW/KW \\		
V3   & 00 50 27.76 +85 15 09.1 & 15.66(1)  & 1.05(5) & $1214_{-34}^{+43}$ &0.80(3)   &-0.09(4)   & -1.88(3)  &0.72(5) & 0.28575(2)  &  no  & $5391 \pm 155$  & EW/KW   \\
V4   & 00 50 50.37 +85 16 12.9 & 15.688(9) & 1.06(4) & $2077_{-109}^{+87}$ &0.47(3)   &-2.44(3)   & -0.79(3)  &0.46(4) & 0.34209(2)  &  yes & $5502 \pm 340$  & EW/KW  \\
V5   & 00 46 12.12 +85 14 02.0 & 16.14(1)  & 1.10(5) & $	1713_{-86}^{+101}$ &0.54(3)   &-1.87(4)   & -0.94(4)  &0.44(9) & 0.32849(4)  & yes  & $5464 \pm 270$  & EW/KW   \\
V6   & 00 47 16.44 +85 15 35.4 & 15.933(3) & 1.04(1) & $1758_{-88}^{+78}$ &0.54(3)   &-2.24(4)   & -0.98(3)  &0.14(4) & 0.33014(6)  &  yes & $5447 \pm 231$  & EW/KW   \\    
V7  & 00 33 48.89 +85 29 22.3 & 15.147(7) & 1.01(3) & $1105_{-25}^{+22}$  &0.88(2)   &6.05(2)    & 4.07(3)   &0.45(4) & 0.31602(2)  &  no  & $5552 \pm 223$  & EW      \\
V8  & 00 44 10.18 +84 54 13.0 & 17.480(8) & 1.84(5) & $948_{-46}^{+49}$  &0.96(7)   &-4.04(8)   & -1.57(8)  &0.8(1)  & 0.27707(4)  & no   & $4252 \pm 315$  & EW     \\
V9  & 00 49 22.30 +84 52 57.6 & 16.392(7) & 1.01(2) & $2990_{-286}^{+344}$ &0.31(4)   &-1.38(5)   & -0.61(4)  &0.43(5) & 0.38597(5)  &  yes  & $5578 \pm 166$  & EW     \\
V10  & 00 51 15.03 +85 24 51.1 & 15.512(3) & 0.97(1) & $1716_{-68}^{+92}$  &0.55(2)   &-2.24(3)   & -0.60(3)  &0.14(3) & 0.35794(5)  &  yes & $5648 \pm 119$  & EW      \\
V11  & 01 01 50.68 +85 24 00.3 & 13.005(4) & 0.793(8) & $1029_{-202}^{+271}$ &0.6(3)    &-9.4(3)	& 8.4(3)	&0.11(4) & 0.321(3)    &  no  & --                 & EW      \\
V12   & 00 48 22.88 +85 15 54.9 & 15.826(5) & 1.21(2)  & $2054_{-72}^{+98}$ &0.49(3)   &-2.46(4)   & -1.06(3)  &0.31(4) & 0.58426(8)  &  yes & $5287 \pm 458$      & EB      \\
V13  & 00 52 08.76 +85 19 05.9 & 17.622(4) & 1.34(2) & $1669_{-128}^{+183}$  &0.52(7)   &-2.3(1)    & -0.86(8)  &0.2(1)  & 0.3048(3)   &  yes & $4842 \pm 358$      & EB    \\
V14  & 01 02 23.28 +85 23 49.1 & 13.49(1)  & 0.64(5)  & $3705_{-131}^{+143}$ &0.24(1)   &8.21(1)	& -0.95(1)  &0.47(4) & 0.4980(2)  &  no   & $6673 \pm 228$      & RRAB   \\
V15  & 00 34 05.31 +84 51 59.0 & 14.106(6) & 0.72(1)  & $1416_{-261}^{+452}$  &-1.2(6)   &3.9(7)	    & 2.4(8)	&0.14(4) & 0.23646(9)  &  no  &  --                 & RRC    \\
V16  & 00 57 47.95 +85 02 29.0 & 13.540(2) & 0.889(6)  & $773_{-6}^{+7}$ &1.27(1)   &-13.17(1)  & 7.97(2)   &0.16(4) & 0.873(2)    &  no  & $5673 \pm 129$  & VAR    \\
V17  & 00 43 23.96 +85 20 32.5 & 14.986(3) & 0.762(7)  & $1863_{-48}^{+49}$ &0.51(2)   &-2.36(2)   & -1.03(2)  &0.05(3) & 0.227(1)    &  yes & $6246 \pm 246$      & VAR    \\
V18  & 00 48 54.43 +84 58 31.6 & 15.492(3) & 0.803(8)  & $1850_{-76}^{+93}$ &0.51(2)   &5.33(3)    & 3.93(3)	&0.09(4) & 0.31698(9)  &  no  & $6116 \pm 179$      & EW \\
V19  & 00 52 37.72 +85 10 34.6 & 14.599(3) & 0.888(6)  & $1841_{-49}^{+84}$ &0.51(2)   &-2.27(2)   & -0.82(2)  &0.62(6) &--  &  yes & $5888 \pm 284$      & EA     \\
V20  & 00 59 34.13 +85 03 08.4 & 12.802(3) & 0.835(5) & $635_{-5}^{+6}$  &1.55(1)   &8.11(2)    & -5.79(2)  &0.48(5) & --       &  no  & $5859 \pm 105$      & EA     \\
V21  & 00 44 52.20 +85 15 54.3 & 15.596(3) & 0.969(8)  & $1831_{-102}^{+96}$ &0.52(3)   &-2.24(3)   & -1.21(3)  &0.51(6) & -- &  yes & $5644 \pm 128$      &EA \\
V22  & 00 32 21.20 +85 18 38.0 & 14.341(4) & 1.36(2)  & $1825_{-43}^{+50}$ &0.52(1)   &-2.24(2)   & -1.17(2)  &0.13(3) &  --      &  yes & $4703 \pm 143$      & L:     \\
V23  & 00 39 00.88 +85 28 15.4 & 13.319(4) & 1.33(1)  & $3211_{-125}^{+108}$ &0.28(1)   &-3.20(1)   & 1.29(1)   &0.10(3) & -- &  no  & $4822 \pm 231$      & L:     \\	
V24  & 00 42 40.49 +85 16 49.4 & 15.005(3) & 1.196(9)  & $1821_{-65}^{+77}$ &0.52(2)   &-2.41(2)   & -1.03(2)  &0.10(3) &-- &  yes & $5163 \pm 316$      & LB     \\
V25  & 00 45 23.00 +85 12 38.0 & 13.770(4) & 1.17(2)  & $1906_{-47}^{+42}$ &0.50(1)   &-2.30(1)   & -0.98(1)  &0.23(3) & -- & yes & $5169 \pm 296$      & RS:     \\

\hline
\end{tabular}
\label{tab:vari}
\footnotesize{\textbf{Notes} The classification of variable stars follows the abbreviated variable star types of VSX catalog. EW/KW indicate contact systems of W Ursae Majoris-type eclipsing variables; EB indicate $\beta$ Lyrae-type eclipsing systems; EA indicate $\beta$ Persei-type (Algol) eclipsing systems; RS indicate RS Canum Venaticorum-type binary systems; VAR indicate variable star of unspecified type; RRAB indicate RR Lyrae variables with asymmetric light curves, while RRC indicate RR Lyrae variables with nearly symmetric light curves; L indicate slow irregular variables, while LB indicate slow irregular variables of late spectral types. And the colons after some types indicate that the type of the variable star is not yet determined.}
\end{table*}

\subsection{Cluster membership of the detected variables}
\label{sect:analysis}

To identify the cluster membership of the detected variable stars in this work,we cross-matched our coordinates of variable stars with \citet{2020A&A...640A...1C}, for which provided 857 cluster members with probabilities over 70\% of open cluster NGC 188 using the membership assignment code Unsupervised Photometric Membership Assignment in Stellar Clusters (UPMASK, \cite{2014A&A...561A..57K}) based on the Gaia DR2 database \citep{2018A&A...616A...4E}. Fourteen variables are identified as cluster members with 100\% membership probabilities and the others are possible field stars for their larger proper motions in declination direction compared with cluster members. 

\begin{figure*}
   \centering
   \includegraphics[width=\linewidth]{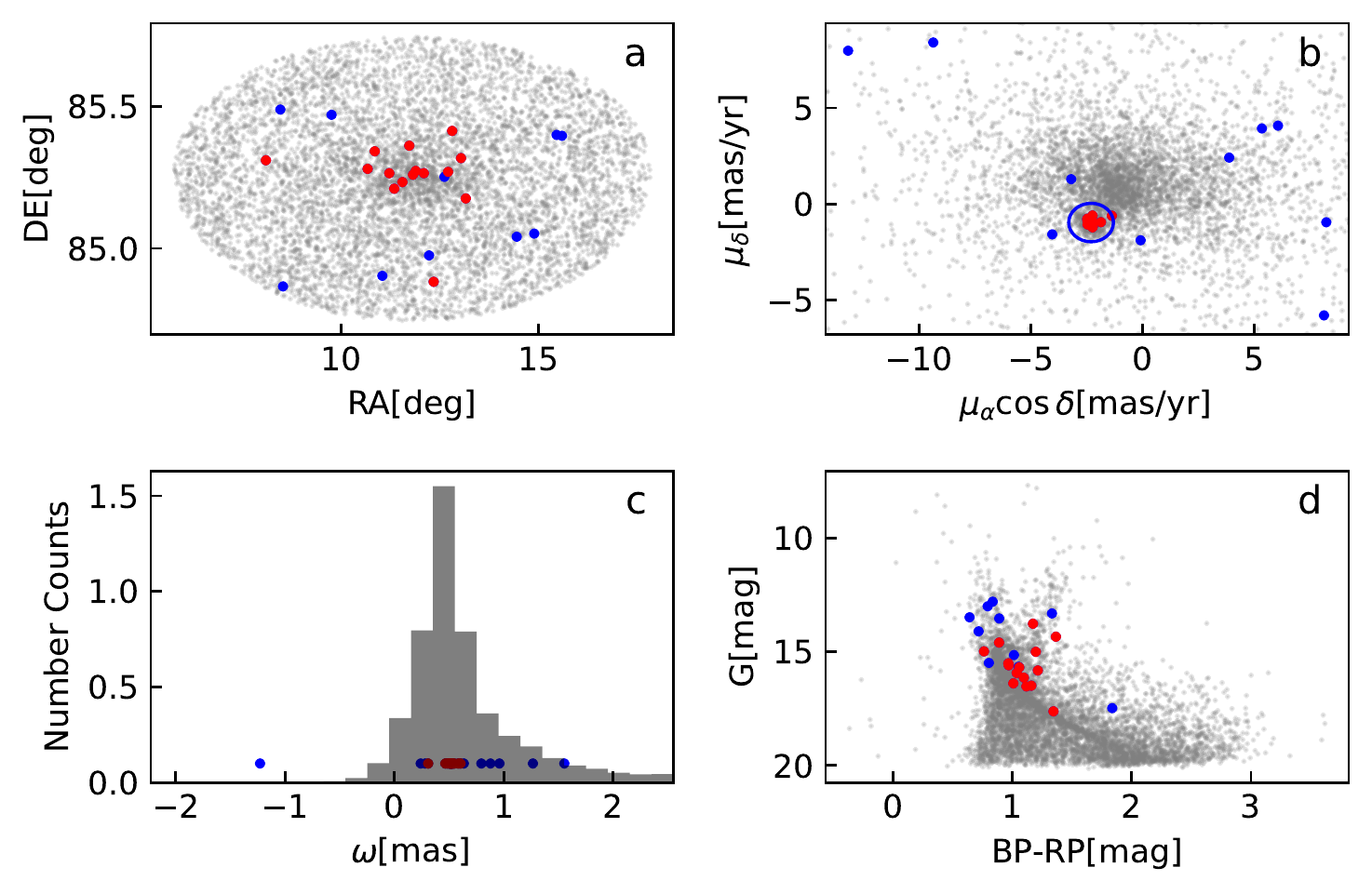}
   \caption{ (a) spatial distribution for the stars in the field of NGC 188 and 25 variable stars; (b) proper motion distribution; (c): histogram of parallax ($\omega$); (d) observed color-magnitude diagram without reddening considered. In the panels, light grey dots represent the Basic Sources of Gaia DR3. Red and blue dots represent the variable members and non-members, respectively.}
   \label{fig:gaia}
\end{figure*}

\cite{2022A&A...659A..59T} points out that \citet{2018A&A...618A..93C} might ignore the cluster members in the peripheral regions of OCs. In order not to miss the cluster member variables, we took advantage of Gaia DR3 \citep{{2022yCat.1355....0G}}, which provides more exquisite astrometric precision than Gaia DR2, to revisit the memberships of the detected variable stars. First, we queried a cone of 30 arcmin radius around the cluster centre with non-zero astrometric and photometric parameters, as well as errors in G mag smaller than 0.005, to create Basic Sources. Figure \ref{fig:gaia} (a) presents the spatial positions of the Basic Sources and the 25 variable stars. The Basic Sources and variables are represented in grey and red/blue dots. The cluster members are obviously concentrated in the space center. In proper motion space, as shown in Figure \ref{fig:gaia} (b), we set a blue circular region centered on $(pmRA, pmDE) = (- 2.307, - 0.960)~mas/yr$ with radius $1.0~mas/yr$ as the selection criteria, and fifteen variable stars are contained in the circle which are labeled in red dots, including all the fourteen variables identified in the preceding paragraph and V9. Then we checked all the stellar parameters of the fifteen stars in the other subgraphs in \ref{fig:gaia}. In Figure \ref{fig:gaia} (a), most red dots are concentrated in the center of the subgraph, V22 and V9 are two slightly distant dots. Figure \ref{fig:gaia} (c) is the histogram of parallax($\omega$) and Figure \ref{fig:gaia} (d) presents the observed color-magnitude diagram without reddening considered. None of the fifteen stars can be excluded as non-members. To confirm the membership of V9, we checked the catalog of \citet{2018A&A...618A..93C}, which provided the membership probabilities for 883 sources in the field of NGC 188 using UPMASK, and no records found. However, we found that \cite{2003AJ....126.2922P} classified V9 as a possible member star ( the probability is 73 \% ) using the astrometry of Tycho-2 catalog.     

\begin{figure*}
\centering
\includegraphics[width=0.6\columnwidth]{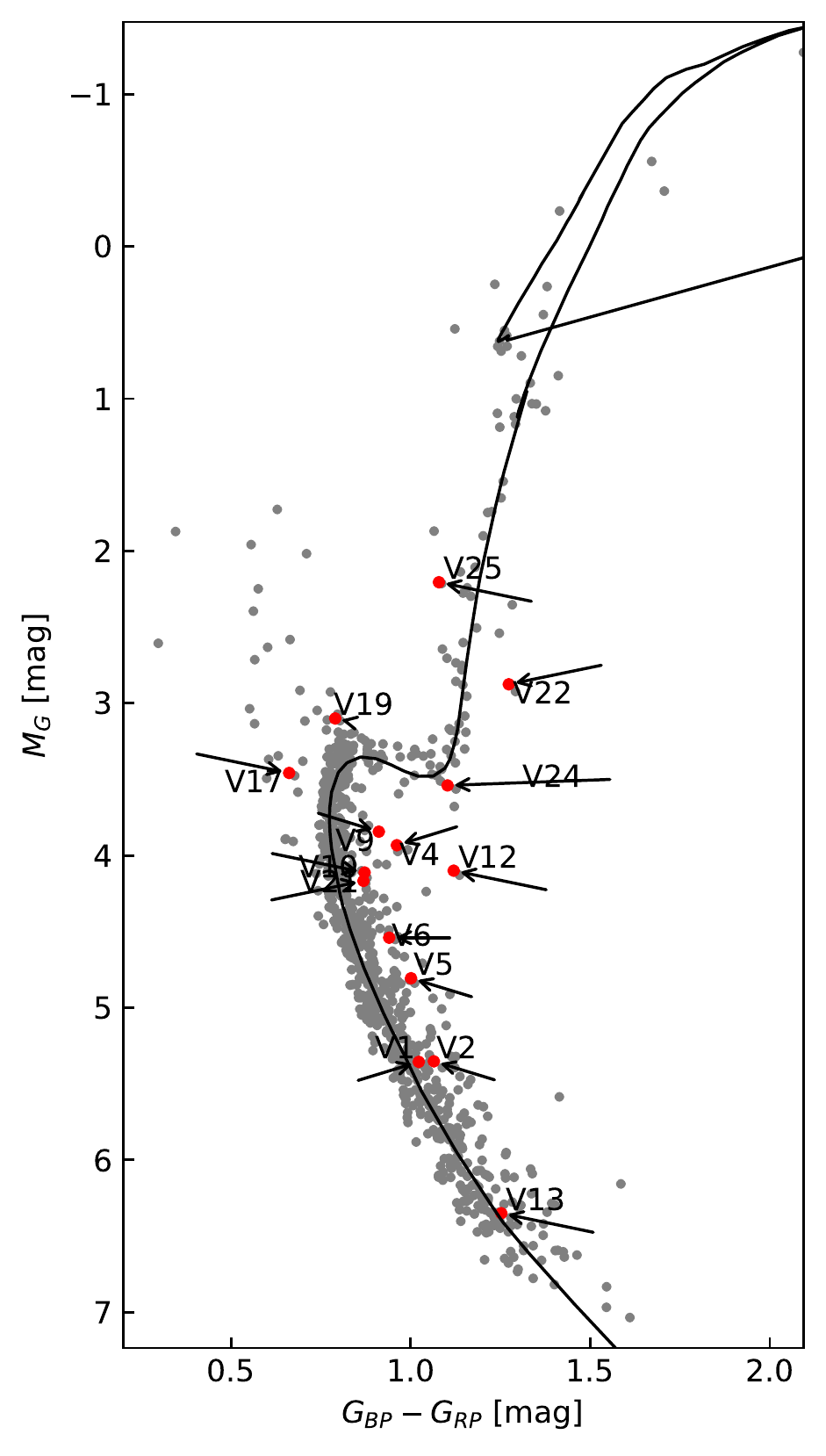}
\caption{ CMD for cluster members of NGC 188 with the positions of the 16 cluster variables marked in red color. The cluster members are provided by CG20 with probabilities over 70 \%. The cluster parameters used in the Padova theoretical isochrone fitting are come from CG20 and \cite{2022MNRAS.513.5387S}.}
\label{fig:cmd}
\end{figure*}

The absolute magnitude $M_G$ versus intrinsic color item $G_{BP} - G_{RP}$ CMD for the open cluster NGC 188 is shown in Figure \ref{fig:cmd}. The absolute magnitude $M_G$ is transformed by the observational magnitude Gmag and distance of each member star. The distances are taken from \cite{2021AJ....161..147B} estimated by probabilistic methods using a 3D a priori model of the Galaxy based on Gaia EDR3 data. We calculated the extinction coefficients $A_G, A_{BP}$, and $A_{RP}$ as follows:

\begin{equation}\label{eq1}
  A_M/A_V=c_{1M}+c_{2M}(G_{BP}-G_{RP})+c_{3M}(G_{BP}-G_{RP})^2+c_{4M}(G_{BP}-G_{RP})^3+c_{5M}A_V+c_{6M}{A_V}^2+c_{7M}(G_{BP}-G_{RP})A_V
\end{equation}

This is the transformation relation for Gaia bands defined by \citet{2018A&A...616A..10G}, where M represents for G, BP or RP band, and $c_{1...7M}$ represent a set of coefficients. 

The Padova theoretical isochrone \citep{2012MNRAS.427..127B} is represented by a black solid line in Figure \ref{fig:cmd}. For the metallicity, we adopted the metal value $[Fe/H] = + 0.064 \pm 0.018$ dex provided by WIYN/Hydra spectra of \cite{2022MNRAS.513.5387S}. Other cluster parameters  adopted are taken from  \cite{2020A&A...640A...1C} (log(t) = 9.85 yr, Z = 0.0152, V - $M_{V}$ = 11.15 mag, $A_{V}$ = 0.21 mag).

\section{Results and Discussion}
\label{sect:discussion}
\subsection{Compared to Previous Works}\label{sub:comp}

The four well-known W UMa variable stars identified by \cite{1964AN....288...49H}, V1 - V4, are easily detected in our observations. The suspected variable NSV 395 was not included in our field of view. All the variables identified by \cite{1987ApJ...314..585K} were detected in our observations, but only the light curves of three variables(V5,V6, V12), showed large enough variabilities. NSV 15158 (V25) was confirmed in this work. The variations of the variables discovered by \cite{2002ChJAA...2..481X} were also  detected in this work, except two outside of the field of view. None of the faint variables found by \cite{2003AJ....126..276K} were confirmed in our catalog. Among the eight variables identified by \cite{2004MNRAS.355.1369Z}, three of them were outside of the field of view, and no variations were found in the other three light curves, only two variables(V9, V15) were confirmed in this work. It is difficult to observe the variations of the BY Draconis type variables discovered by \cite{2008AcA....58..263M} due to the telescope's limitation. The eclipsing binary (V13) and one BY Draconis type star (V21) of \cite{2008AcA....58..263M} were detected in this work. Among the eighteen variables of \cite{2013OEJV..157....1P}, eleven were out of the field of view, and the light curves of two variables were flat in our observations, the other five were confirmed in this work. The four suspected variables, are detected in our observation, while NSV 15164 is saturated, and the light curves of the other three variables did not show changes. None of the variables discovered by the TAROT Suspected Variable Star Catalog (TSVSC1) are confirmed by this work. For the two variables detected by ASAS-SN, the detached eclipsing binary (V20) was detected, but there were no variations detected in the light curve of another variable in this work. All the detected 24 known variables are listed in Table \ref{tab:vari}. Among the 24 detected known variables in the field of NGC 188, good-phased light curves were recorded for the first 17 variables listed in Table \ref{tab:vari} as shown in Figure \ref{fig:fig1}. As shown in Figure \ref{fig:fig2}, the observational durations of the other seven variables are not long enough to determine the period of these variable stars. All the memberships of the detected variables are listed in Table \ref{tab:vari}. We didn’t investigate the known variable stars if our data are basically consistent with the previous conclusions.

V16 and V17 were found as variable stars by \citet{2013OEJV..157....1P}. Because of the low amplitudes of their light variations, the classifications for the two stars still remain unknown. V16 is a certain foreground field star and is a spectroscopic variable star classified by \cite{2020ApJS..249...22T} based on the data of LAMOST DR4. V17 is a certain blue straggler star (BSS) based on the location of the CMD, and it was classified as a single-lined BSS with rapid rotation by \citet{2009AJ....137.3743G}.

V21 was classified as a BY Draconis type variable by \citet{2008AcA....58..263M}. It was detected by our observations because an about 0.5 mag eclipse appeared in the light curve of this star, which implies that it might be an Algol-type eclipsing binary. The location of this star in our abosute $M_G$ versus $G_{BP}-G_{RP}$ CMD also reinforces this view. More observations for this star are needed to determine its period and properties.

\subsection{Classification of New Variables}\label{sub:new}

One new periodic variable star (V18) is identified in this work. The phased light curves are shown in Figure \ref{fig:fig1}, and the basic parameters for the variable are listed in Table \ref{tab:vari}. As discussed above, this is a field star with definite variations of brightness. Based on the distances of the star, V18 is a field star mixed with cluster members. We considered the period, amplitude, light-change curve shape, effective temperature, and positions on the CMD to make the classification of the variable stars obtained in this study.  We checked the star's position on the CMD and compared with the statistical positions of the periodic variable stars on the CMDs given by \citep{2019A&A...623A.110G}, and found it could be a W UMa binary.

\section{Conclusions}
\label{sect:conclusion}
In this paper, we have presented the time-series V-band photometric survey of the open cluster NGC 188, with particular emphasis on variable stars. The results of this study are the following:

i). We detected 25 variable stars in a $55 \times 45~arcmin^2$ field of view around the cluster, including one new variable. Their memberships are determined by the research of CG20 and reconfirmed by Gaia DR3. Most results are consistent with CG20, except V9, for which is a possible cluster member in our research. Our results suggest that 15 variables are cluster members while the other 14 stars belong to the field star population.

ii). Based on the behaviors and periods of the light curves as well as their positions on the CMD,
we discussed the classifications of the 25 variable stars. Most results of the known variables are coincident with the VSX catalog, except V21 (V0769 Cep), which is preferred to be an EA-type eclipsing binary than a BY Draconis type variable star. The new variable is likely to be a W UMa eclipsing star. The detection and analysis of the variable stars in the old open cluster NGC 188 yield valuable samples, especially for the study of W UMa stars.

\begin{acknowledgements}
We are grateful to an anonymous referee for valuable comments which have improved the paper significantly. This work has been financially supported by the Resource sharing platform construction project of Xinjiang Uygur Autonomous Region (No. PT2306) and the Chinese Academy of Sciences (CAS) "Light of West China" Program (No. 2020-XBQNXZ-016). This work has made use of data from the European Space Agency (ESA) mission {\it Gaia} (\url{https://www.cosmos.esa.int/gaia}), processed by the {\it Gaia} Data Processing and Analysis Consortium (DPAC, \url{https://www.cosmos.esa.int/web/gaia/dpac/consortium}). The CCD photometric data of NGC 188 were obtained with the Nanshan 1 m telescope administered by Xinjiang Astronomical Observatory.
\end{acknowledgements}

\label{lastpage}

\end{document}